\documentclass{article}

\usepackage{graphicx,amsfonts,amsmath}
\usepackage{booktabs}
\usepackage{algorithm}
\usepackage{xspace}
\newcommand{\gemm}{\texttt{gemm}\xspace}
\newcommand{\fgemm}{\texttt{fgemm}\xspace}
\newcommand{\pfgemm}{\texttt{pfgemm}\xspace}
\newcommand{\dgemm}{\texttt{dgemm}\xspace}
\newcommand{\trsm}{\texttt{trsm}\xspace}
\newcommand{\utrsm}{\texttt{utrsm}\xspace}
\newcommand{\ftrsm}{\texttt{ftrsm}\xspace}
\newcommand{\pluq}{\texttt{pluq}\xspace}
\newcommand{\pftrsm}{\texttt{ftrsm}\xspace}
\newcommand{\laswp}{\texttt{laswp}\xspace}
\newcommand{\flaswp}{\texttt{flaswp}\xspace}
\newcommand{\pflaswp}{\texttt{flaswp}\xspace}
\newcommand{\getrf}{\texttt{getrf}\xspace}

\newcommand{\dgetrftile}{\texttt{getrf\_tile}\xspace}
\newcommand{\fgetrf}{\texttt{fgetrf}\xspace}
\newcommand{\libkomp}{\texttt{libkomp}\xspace}
\newcommand{\libgomp}{\texttt{libgomp}\xspace}
\newcommand{\xkaapi}{\texttt{XKaapi}\xspace}
\newcommand{\fflasffpack}{\texttt{fflas-ffpack}\xspace}
\newcommand{\plasmaquark}{\texttt{Plasma-Quark}\xspace}
\newcommand{\Z}{\ensuremath{\mathbb{Z}}\xspace}
\usepackage[noend]{algpseudocode}
\algrenewcommand\algorithmicrequire{\textbf{Input:}}
\algrenewcommand\algorithmicensure{\textbf{Output:}}
\algrenewcommand\algorithmicreturn{\textbf{Return}}
\algrenewcommand\Return{\State\algorithmicreturn{} } 

\usepackage{multirow,rotating}

\author{
Jean-Guillaume Dumas\footnote{
Universit\'e de Grenoble. 
Laboratoire LJK,
umr CNRS, INRIA, UJF, UPMF, GINP.
51, av. des Math\'ematiques, F38041 Grenoble, France.
}~\footnotemark[4], 
\and Thierry Gautier\footnote{
INRIA.
Laboratoire LIG,
umr CNRS, INRIA, UJF, UPMF, GINP.
51, av. J. Kuntzmann, F38330 Montbonnot St-Martin, France.
}~\footnotemark[4]
\and Cl\'ement Pernet\footnote{
Universit\'e de Grenoble.
Laboratoire de l'Informatique du Parall\'elisme 
umr CNRS, INRIA, UCBL, \'ENS de Lyon.
46 All\'ee d'Italie, F69364 LYON Cedex 07, France.
}~\footnotemark[4]
\and Ziad Sultan\footnotemark[1]~\footnotemark[2]
\footnote{
\href{mailto:Jean-Guillaume.Dumas@imag.fr}{Jean-Guillaume.Dumas@imag.fr},
\href{mailto:Thierry.Gautier@imag.fr}{Thierry.Gautier@imag.fr},
\href{mailto:Clement.Pernet@imag.fr}{Clement.Pernet@imag.fr},
\href{mailto:Ziad.Sultan@imag.fr}{Ziad.Sultan@imag.fr}.
}
}

\title{Parallel computation of echelon forms\thanks{\small This work is partly funded by the HPAC project of the French Agence Nationale de la Recherche (ANR~11~BS02~013).}}

\date{}

\usepackage{color,svgcolor}
\usepackage[plainpages=true,hypertexnames=false]{hyperref}
\makeatletter
\hypersetup{
pdftitle={\@title},
pdfauthor={J-G. Dumas, T. Gautier, C. Pernet, Z. Sultan},
breaklinks=true, 
colorlinks=true,
 linkcolor=darkred,
 citecolor=blue,
 urlcolor=darkgreen,
}
\makeatother

\begin{document}
\maketitle
\begin{abstract}
We propose efficient parallel algorithms and implementations on
shared memory architectures of LU factorization over a finite field.
Compared to the corresponding numerical routines, we have identified three main
difficulties specific to linear algebra over finite fields.
First, the arithmetic complexity could be dominated
by modular reductions. Therefore, it is mandatory to delay as much as
possible these reductions while mixing fine-grain parallelizations of
tiled iterative and recursive algorithms.
Second, fast linear algebra variants, e.g., using Strassen-Winograd
algorithm, never suffer from instability and can thus be widely used in
cascade with the classical algorithms. There, trade-offs are to be
made between size of blocks well suited to those fast variants or to
load and communication balancing. 
Third, many applications over finite fields require the rank profile
of the matrix (quite often rank deficient) rather than the solution to
a linear system. It is thus important to design parallel algorithms that
preserve and compute this rank profile. Moreover, as the rank profile is only
discovered during the algorithm, block size has then to be dynamic.
We propose and compare several block decomposition: tile iterative with
left-looking, right-looking and Crout variants, slab and tile recursive.

Experiments demonstrate that the tile recursive variant performs better and
matches the performance of reference numerical software when no rank deficiency
occur. Furthermore, even in
the most heterogeneous case, namely when all pivot blocks are rank
deficient, we show that it is possbile to maintain a high efficiency.
\end{abstract}

\section{Introduction}

Triangular matrix factorization is a main building block in  computational  linear
algebra. 
 
Driven by a large range of applications in computational sciences, parallel
numerical dense LU factorization has been intensively studied since 
several decades which results in software of great maturity (e.g., LINPACK is used for benchmarking the efficiency of the top
500 supercomputers~\cite{DLP03}).
 
More recently, efficient sequential exact linear algebra
routines were developed~\cite{DGP08}. They are used in algebraic cryptanalysis, computational number theory, or integer linear programming and they benefit from the experience in numerical linear
algebra. In particular, a key point there is to embed the finite field elements
in integers stored as floating point numbers, and then rely  on the efficiency of the
floating point matrix multiplication \texttt{dgemm} of the BLAS. The conversion
back to the finite field, done by costly modular reductions, is delayed as much as possible.
 
Hence a natural ingredient in the design of efficient dense linear algebra routines is the
use of block algorithms that result in gathering arithmetic operations in
matrix-matrix multiplications. Those can take full advantage of vector instructions and has a high computation
per memory access rate, allowing to fully overlap the data accesses by
computations and hence delivers peak performance efficiency.
 
In order to exploit the power of multi-core and many-core architectures, we now
investigate the parallelization of the finite field linear algebra routines. 
We report in this paper the conclusions of our experience in parallelizing
exact LU decomposition for shared memory parallel computers. We try to emphasize
which specificities of exact computation domains led us to use different
approaches than that of numerical linear algebra.
 
In short, we will illustrate that numerical and exact LU factorization mainly differ in the 
following aspects:
\vspace{-3mm}
\begin{itemize}
\item the pivoting strategies,
\item the cost of the arithmetic (of scalars and matrices),
\item the treatment of rank deficiencies.
\end{itemize}
\vspace{-2mm}
Those have a direct impact on the shape and granularity of the block
decomposition of the matrix used in the computation.

\vspace{-1em}
\paragraph{Types of block algorithms.}
Several schemes are used to design block linear algebra algorithms: the splitting can occur on one
dimension only, producing row or column slabs~\cite{KlvdGe95}, or both
dimensions, producing tiles~\cite{BLKD07}. Note that, here, we denote by tiles a
partition of the matrix into sub-matrices in the mathematical sense regardless
what the underlying data storage is.

\begin{figure}[ht]
\begin{center}
  \includegraphics[width=.18\textwidth]{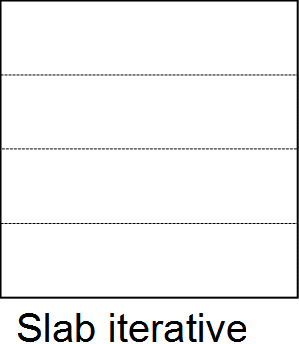}
\hfill
  \includegraphics[width=.18\textwidth]{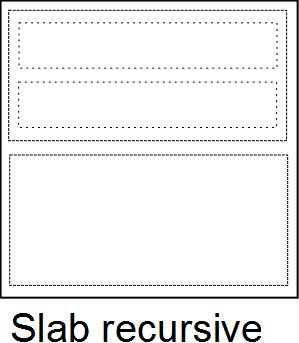}
\hfill
  \includegraphics[width=.18\textwidth]{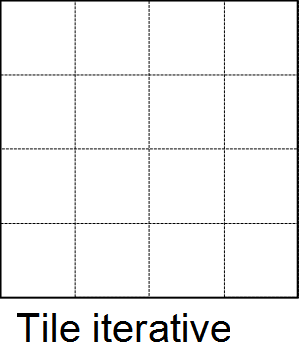}
\hfill
  \includegraphics[width=.18\textwidth]{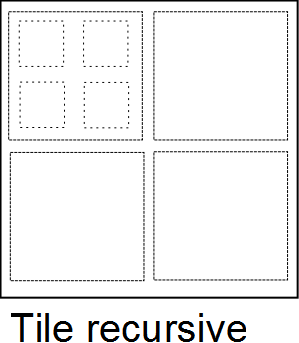}\vspace{-5pt}
\end{center}
  \caption{Main types of block splitting}
  \label{fig:blockalg}\vspace{-5pt}  
\end{figure}

Algorithms processing blocks can be either iterative or recursive.
Figure~\ref{fig:blockalg} summarizes some of the various existing block
splitting obtained by combining these two aspects.
Most numerical dense Gaussian elimination algorithms, like in \cite{BLKD07}, use
tiled iterative block algorithms. In~\cite{DFLL11} the classic tiled iterative
algorithm is combined with a slab recursive one for the panel elimination.
Over exact domains, recursive algorithms are preferred to benefit from fast
matrix arithmetic (see below).
 
Slab recursive exact algorithms can be found in~\cite{JPS13} and references therein
and~\cite{DPS13} presents a tiled recursive algorithm.
 
\vspace{-1em}
\paragraph{The granularity} is the block dimension (or the dimension of the
smallest blocks in recursive splittings). 
Matrices with dimensions below this threshold are treated by a
base-case variant (often referred to as the panel factorization).  
It is an important parameter for optimizing efficiency: a finer grain allows
more flexibility in the scheduling when running numerous cores, but it also
challenges the efficiency of the scheduler and can increase the bus traffic.

\vspace{-1em}
\paragraph{The cost of the arithmetic.} In numerical linear algebra, the cost of
arithmetic operations is more or less associative: with dimensions above a rather low
threshold (typically a few hundreds), sequential matrix multiplication of the
BLAS reaches the peak efficiency of the processor. Hence the granularity has
very little impact on the efficiency of a block algorithm run in sequential.
On the contrary, over a finite field, a small granularity can imply a larger
number of costly modular reductions, as we will show in Section~\ref{sec:modcount}. 
Moreover,  numerical stability is not an issue over a finite field, and asymptotically fast matrix
multiplication algorithms, like Winograd's variant of Strassen
algorithm~\cite[\S 12]{GG99} can be used on top of the BLAS.  
Their speed-up increases with matrix dimension. The cost of
sequential matrix multiplication over finite field is therefore not associative:
a larger granularity delivers better sequential efficiency.

\paragraph{Pivoting strategies and rank deficiencies.}
In dense numerical linear algebra, pivoting is used to ensure good
numerical stability and good data locality~\cite{GoVa96}.  
In the context of dense {\em exact} linear algebra, stability is no longer an 
issue. Instead, only certain pivoting strategies  will reveal 
the echelon form or, equivalently, the rank profile of the matrix~\cite{JPS13,DPS13}. This is a key
invariant used in many applications using exact Gaussian elimination, such as
Gr\"obner basis computations~\cite{F99a} and computational number
theory~\cite{stein2007modular}.

Over exact domains, the rank deficiency of some blocks also leads to unpredictable
dimensions  of the tiles or slabs, as will be illustrated in Section~\ref{sec:rankdef}.
This makes the block splitting necessarily dynamic contrarily to the case of
numerical LU factorization where all panel blocks usually have full rank
and the splitting is done statically according to a granularity parameter.

Consequently the design of a parallel exact matrix factorization 
 necessarily differs from the numerical algorithms as follows:
\vspace{-3mm}
\begin{itemize}
\item granularity should be as large as possible, to reduce modular reductions
  and benefit from fast matrix multiplication;
\item algorithms should preferably be recursive, to group arithmetic operations
  in matrix products as large as possible.
\item block splitting and pivoting strategies must preserve and reveal the rank
  profile of the matrix
\end{itemize}
\vspace{-2mm}
It also implies several requirements on the parallel run-time being used:
\vspace{-3mm}
\begin{itemize}
\item the block splitting has to be dynamically computed;
\item the computing load for each task is not known in advance (some panel
  blocks may have high rank deficiency), making the tasks very heterogeneous.
\end{itemize}
\vspace{-2mm}

This motivated us to look into parallel execution runtimes using tasks with work-stealing based
scheduling.

All experiments have been conducted on a 32 cores Intel Xeon E5-4620 2.2Ghz
(Sandy Bridge) with L3 cache(16384 KB). The numerical BLAS is ATLAS v3.11.4,
LAPACK v3.4.2 and PLASMA v2.5.0. We used X-KAAPI-2.1 version with last git commit:
xkaapi\_2.1-30-g263c19c638788249. The gcc compiler version used is gcc 4.8.2 that
supports OpenMP 3.1.

We introduce in Section~\ref{sec:prelim} the algorithmic building blocks on
which our algorithms will rely and the parallel programming models and runtimes
that we used in our experiments.
In order to handle each problem separately, we focus in
Section~\ref{sec:fullrank} on the simpler case where no rank deficiency
occur. In particular Section~\ref{sec:modcount} presents detailed analysis of the
number of modular reductions required by various block algorithms including the tiled
and slab recursive, the left-looking, right-looking and Crout variants of the
tiled iterative algorithm. Lastly Section~\ref{sec:rankdef}  deals with elimination
with rank deficiencies. We there present and compare new slab iterative,
tiled iterative and tiled recursive parallel algorithms that preserve
rank profiles. We then show that the latter can match state of the art
numerical routines, even when taking rank deficiencies into account.

\section{Preliminaries}
 
\label{sec:prelim}

\subsection{Auxiliary sequential routines}

All block algorithms that we will describe rely on four type of operations that
we denote using the BLAS/LAPACK naming convention:
\vspace{-2mm}
\begin{description}
  \item[\gemm:] general matrix multiplication, computing $C\leftarrow \alpha
    A\times B +\beta C$,
  \item[\trsm:] solving upper/lower triang. syst. with matrix right/left h.s $B\leftarrow B
    U^{-1}$.
  \item[\laswp:] permuting rows or columns by sequence of swaps. 
 
  \item[\getrf:] 
 
computing $(P,L,U,Q)$, $L$ and $U$ stored in place
    of $A$, s.t. $A=PLUQ$.
\end{description}
\vspace{-2mm}

A first prefix letter \texttt{d} or \texttt{f} is specifies if the routine
works over double precision floating point numbers or finite field coefficients
and an optional prefix \texttt{p} stands for parallel implementation.
Our implementations use the sequential routines of the \fflasffpack
library\footnote{\url{http://linalg.org/projects/fflas-ffpack}}~\cite{DGP08}. There,
the elements of a finite $\Z/p\Z$ for a prime $p$ of size about 20 bits
are integers stored in a double precision floating point number. The sequential
\fgemm routine combines recursive steps of Winograd's algorithm calls to
numerical BLAS \dgemm and reductions modulo $p$ when necessary. The \ftrsm and
\fgetrf routines use block recursive algorithms to reduce most arithmetic
operations to \fgemm. More precisely \fgetrf is either done by a slab
recursive algorithm~\cite{DGP08} or a tile recursive algorithm~\cite{DPS13}.

\subsection{Parallel programming models}

We base our implementation on the task based parallel features of OpenMP
standard.
 
This is motivated by the use of recursive algorithms where tasks are
mandatory. Now in tile iterative algorithms, loop with tasks happen to perform
at least as good as parallel loops.

\libgomp is the GNU implementation of the OpenMP API for multi-platform
shared-memory parallel programming in C/C++ and Fortran.

Alternatively, we also used
\libkomp~\cite{BGD12}, an optimized version of \libgomp, based on the \xkaapi runtime,
that reduces the overhead of the OpenMP directives and handles more
efficiently threads creation, synchronization and management.
In the experiments of the next sections, we will compare efficiency of the same
code linked against each of these two libraries.

\subsection{Parallel matrix multiplication}

In the iterative block algorithms, all matrix product tasks are sequential, on
the contrary the recursive block algorithms must call parallel matrix products
\pfgemm, which we describe here. Operation \pfgemm is of the form 
$C \leftarrow \alpha  A \times B+\beta C$.

In order to split the computation into independent tasks, only the row
dimension of $A$ and the column dimension of $B$ only are split. 
The granularity of the split can be chosen in two different ways: either as a
fixed value, or by a ratio of the input dimension (e.g. the total number of
cores).
We chose the second option that maximizes the size of the blocks while ensuring
a large enough number tasks to for the computing ressources. All our experiments
showed that this option performs better than the first one. 
 
When used as a subroutine in a parallel factorization, it will create more tasks
than the number of available cores, but this heuristic happen to be a good
compromise in terms of efficiency.

\begin{figure}[ht!]
\centering
 
\includegraphics[width=.8\textwidth,angle=0.]{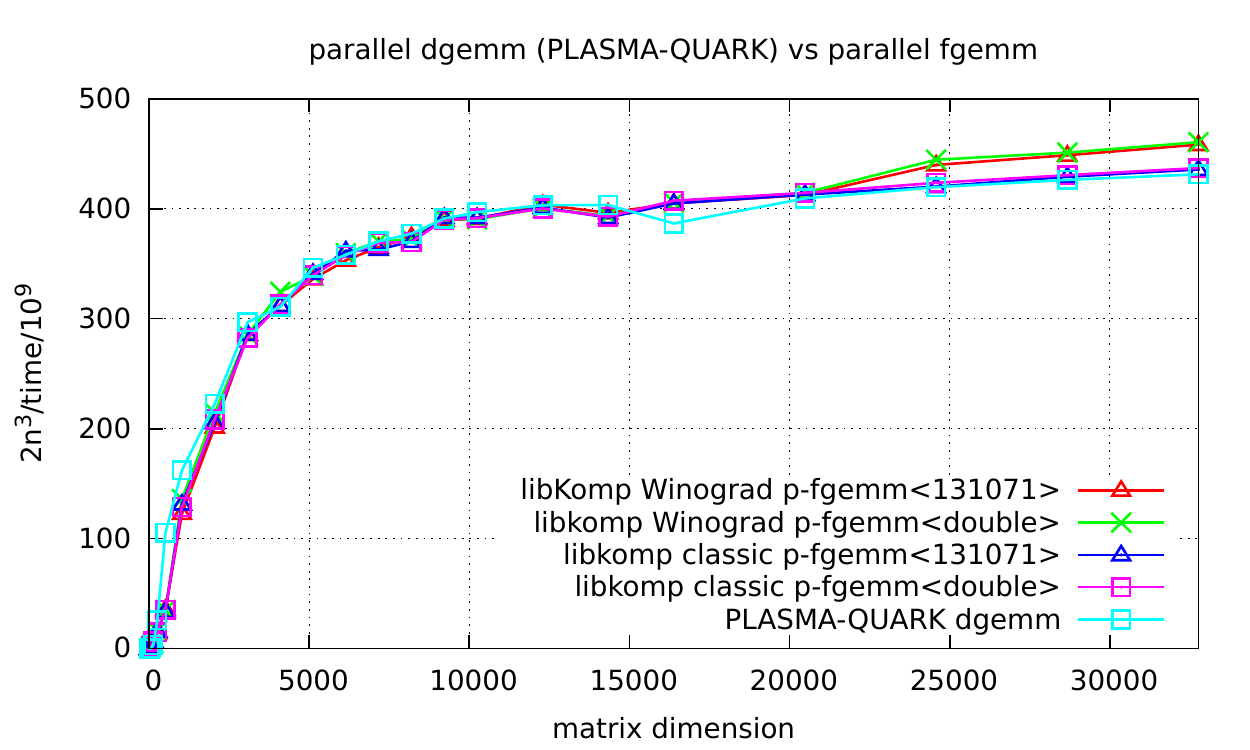}\\

\caption{Comparison of execution time exact vs numeric}
\label{fig:pfgemmtime}\vspace{-5pt}
\end{figure}

Figure~\ref{fig:pfgemmtime} shows the computation time on 32 cores of various
matrix multiplications: the numerical \dgemm implementation of \plasmaquark, 
the implementation of \pfgemm of \fflasffpack using OpenMP tasks, linked
against the \libkomp library. This implementation is run over the finite field
$\Z/131071\Z$ or over field of real double floating point numbers, with or
without fast Strassen-Winograd's matrix product. 
One first notices that most routine perform very similarly. More precisely,
\plasmaquark \dgemm is faster on small matrices but the effect of
Strassen-Winograd's algorithm makes \pfgemm faster on larger matrices, even
on the finite field where additional modular reductions occur.

In terms of speed-up, the \pfgemm reaches a factor of approximately 27 (using
32 cores) whereas the numerical \dgemm of \plasmaquark reaches a factor of 29,
but this is mostly reflect the fact that \dgemm has a less efficient sequential
reference timing since it does not use Strassen-Winograd's algorithm.

Similarly, other basic routines used in the recursive block algorithms, such as
\ftrsm (solving matrix triangular systems) and \flaswp (permuting rows or
columns), have been parallelized by splitting a dimension into a constant number
of blocks (typically the number of cores).

\section{Eliminations with no rank deficiency}
\label{sec:fullrank}

In this section, we make the assumption that no rank deficiency occur during the
elimination of any of the diagonal block. This hypothesis is satisfied by
matrices with generic rank profile (i.e. having all their leading principal
minor non zero). This assumption allows us to focus on the problem of reducing
the modular reduction count.

\subsection{Modular reductions}
\label{sec:modcount}
When computing over finite field, it is of paramount importance to
reduce the number of modular reductions in the course of linear algebra
algorithms. The classical technique is to accumulate several
multiplications before reducing, namely replacing 
$\sum_{i=1}^n (a_ib_i \mod p)$ with $\left(\sum_{i=1}^n a_ib_i\right)$ while keeping
the result exact. If $a_i$ and $b_i$ are integers between $0$ and
$p-1$ this is possible with integer or floating point units if the
result does not overflow, or in other words if
$n(p-1)^2<2^\text{mantissa}$, see, e.g., \cite{DGP08} for more
details. 

This induces a splitting of matrices in blocks of size the largest $n^*$ satisfying
the latter condition. Now the use of block algorithms in parallel, introduces a
second blocking parameter that interferes in the number of reductions.
We will therefore compare the number of modular reductions of three variants of
the tile iterative algorithm (left-looking, right-looking and Crout, see~\cite{DDSV98}), the slab
recursive algorithm of~\cite{DGP08}, and the tile recursive algorithm of~\cite{DPS13}.
For the sake of simplicity, we will always assume that the block dimensions in the
parallel algorithms are always below $n^*$. In other words operations are done with full
delayed reduction for a single multiplication and any number of
additions: operations of the form $\sum a_i b_i$ are reduced modulo
$p$ only once at the end of the addition, but $a \cdot b \cdot c$ requires two
reductions.

For instance, with this model, the number of reductions required by a
classic multiplication of matrices of size $m\times k$ by $k\times n$
is simply: $R_{\gemm}(m,k,n)=mn$.
From \cite[Theorem~3]{DPS13}, this extends also for triangular solving $m\times
m$ to $m\times n$: with
unit diagonal, $R_\utrsm(m,m,n)=mn$ (actually the computation of the lowest row of
the solution requires no modulo as it is just a division by $1$, we will
therefore rather use $R_\utrsm(m,m,n)=(m-1)n$) 
and $R_\trsm(m,m,n)=2mn$ (with the previous refinement for
$R_\utrsm(m,m,n)$, this also reduces to $R_\trsm(m,m,n)=(2m-1)n$).

Table~\ref{tab:blockvariants} sketches the different shapes of the associated
routine calls in the main loop of each variant.
\begin{table}[ht]\center
\begin{tabular}{ccc}
\toprule
Left looking  &
Crout         &
Right looking \\
\midrule
\hspace{-5pt}\begin{minipage}{.33\textwidth}
\begin{algorithmic}
\For{i=1 to n/k}
\State \utrsm((i-1)k,(i-1)k,k)
\State \gemm(n-(i-1)k,(i-1)k,k)
\State \pluq(k,k)
\State \trsm(k,k,n-ik)
\EndFor
\end{algorithmic}
\end{minipage} &
\hspace{-5pt}\begin{minipage}{.33\textwidth}
\begin{algorithmic}
\For{i=1 to n/k}
\State \gemm(n-(i-1)k,(i-1)k,k)
\State \gemm(k,(i-1)k,n-ik)
\State \pluq(k,k)
\State \utrsm(k,k,n-ik)
\State \trsm(k,k,n-ik)
\EndFor
\end{algorithmic}
\end{minipage} &
\begin{minipage}{.31\textwidth}
\begin{algorithmic}
\For{i=1 to n/k}
\State \pluq(k,k)
\State \utrsm(k,k,n-ik)
\State \trsm(k,k,n-ik)
\State \gemm(n-ik,k,n-ik)
\EndFor
\end{algorithmic}
\end{minipage}\\
\bottomrule
\end{tabular}
\vspace{5pt}
\caption{Main loops of the Left looking, Crout and Right looking tile iterative
 block LU factorization (see~\cite{DDSV98})}\label{tab:blockvariants}
\vspace{-5pt}
\end{table}

Then the number of modular reductions required for these different LU
factorization strategies is given in Table~\ref{tab:modcount}.
\begin{table}[ht]
\renewcommand{\arraystretch}{1.5}
\center 
\begin{tabular}{clc}
\toprule
\multirow{3}{*}{\begin{sideways}$k= 1$\end{sideways}} 
& Iterative Right looking & $\frac{1}{3}n^3-\frac{1}{3}n$ \\
& Iterative Left Looking & $\frac{3}{2}n^2-\frac{3}{2}n+1$\\
& Iterative Crout & $\frac{3}{2}n^2-\frac{7}{2}n+3$\\
\midrule
\multirow{3}{*}{\begin{sideways}$k\geq 1$\end{sideways}} 
& Tile Iterative Right looking & $\frac{1}{3k}n^3+\left(1-\frac{1}{k}\right)n^2+\left(\frac{1}{6}k-\frac{5}{2}+\frac{3}{k}\right)n$ \\
& Tile Iterative Left looking & $\left(2-\frac{1}{2k}\right)n^2+\left(-\frac{5}{2}k-1+\frac{2}{k}\right)n+2k^2-2k+1$\\
& Tile Iterative Crout & $\left(\frac{5}{2}-\frac{1}{k}\right)n^2+\left(-2k-\frac{5}{2}+\frac{3}{k}\right)n+k^2$\\
\midrule
& Tiled Recursive & $2n^2-n\log_2 n-n$\\
\midrule
& Slab Recursive & $(1+\frac{1}{4}\log_2 n)n^2-\frac{1}{2}n\log_2 n-n$\\
\bottomrule
\end{tabular}
\vspace{5pt}
\caption{Counting modular reductions in full rank block LU
  factorization of an $n\times n$ matrix modulo $p$ when $np(p-1)<2^\text{mantissa}$,
  for a block size of $k$ dividing $n$.}\label{tab:modcount}
\vspace{-10pt}
\end{table}

The last two rows of the table corresponds
to~\cite[Theorem~4]{DPS13} where $R_\utrsm$ has been refined to $(m-1)n$ as mentioned above.
The first three rows are obtained by setting $k=1$ in the following block
versions. 
The next three rows are obtained via the following analysis
where the base case (i.e. the $k \times k$ factorization) always uses the best
unblocked version, that is the Crout variant described above.
Following Table~\ref{tab:blockvariants}, we thus have:
\begin{itemize}
\item The right looking variant performs $\frac{n}{k}$ such $k\times k$ base cases,
  $\pluq(k,k)$, then, at iteration $i$, $(\frac{n}{k}-i)(\utrsm(k,k,k)+\trsm(k,k,k))$,
  and $(\frac{n}{k}-i)^2$ \gemm(k,k,k), for a total of
  $\frac{n}{k}(\frac{3}{2}n^2-\frac{7}{2}n+3)+\sum_{i=1}^{\frac{n}{k}}
  (n-ik)\left((3k-2) + (\frac{n}{k}-i)k\right)=\frac{1}{3k}n^3+\left(1-\frac{1}{k}\right)n^2+\left(\frac{1}{6}k-\frac{5}{2}+\frac{3}{k}\right)n$.
\item The Crout variant requires:

      at each step, except the first one, to compute $R_{\gemm}(n-ik,ik,k)$
      reductions for the pivot and the elements below and
      $R_\gemm(k,ik,n-(i-1)k)$;
 
      at each step, to perform one base case for the pivot block, to
      solve unitary triangular systems, to the left, below the pivot, using
      $(\frac{n}{k}-i)R_\utrsm(k,k,k)$ reductions and to solve triangular systems to the
      right, using $(\frac{n}{k}-i)R_\trsm(k,k,k) $ reductions. 
 
\item Similarly, the Left looking variant requires $R_\gemm(n-ik, ik,
  k)+R_\pluq(k)+R_\utrsm(ik, ik,k)+R_\trsm(k,k,n-ik)$ reductions in the main loop.
\end{itemize}

In table~\ref{tab:overhead} we see that the left looking variant always performs less modular reductions. 
Then the tiled recursive performs less modular reductions than the Crout variant
as soon as $2\leq k \leq \frac{n}{4}$.

Finally the right looking variant is clearly costs more modular reductions.
 
But the best algorithms here may not perform well in parallel, as will be shown next.

\begin{table}[htbp] 
\begin{center}
 
\renewcommand{\arraystretch}{1.2}
\begin{tabular}{l|ccc|ccc|cc}
\toprule
  & \multicolumn{3}{c}{$k=212$} & \multicolumn{3}{c}{$k=\frac{n}{3}$} & \multicolumn{2}{c}{Recursive}\\
\midrule
   & Right & Crout & Left       & Right & Crout & Left   & Tile & Slab\\
\midrule
n=3000 & \multicolumn{1}{c}{3.02} & 2.10 & \textbf{2.05}  & \multicolumn{1}{c}{2.97} & 2.15 & 2.10  & \multicolumn{1}{c}{2.16} &
2.26\\
\midrule
n=5000 & \multicolumn{1}{c}{11.37} & 8.55 & 8.43  & \multicolumn{1}{c}{9.24} & 8.35 & 8.21  &\multicolumn{1}{c}{\textbf{7.98}} &
8.36\\
\midrule
n=7000 & \multicolumn{1}{c}{29.06} & 22.19 & 21.82  & \multicolumn{1}{c}{22.56} & 22.02 & 21.73  & \multicolumn{1}{c}{\textbf{20.81}} & 21.66\\
\bottomrule
\end{tabular}
\end{center}
\caption{Timings (in seconds) of sequential LU factorization variants on one core}  
\label{tab:overhead} \vspace{-20pt}
\end{table}

\subsection{Parallel experiments}

\begin{figure}[ht!]
\centering
\includegraphics[width=.8\textwidth,angle=0.]{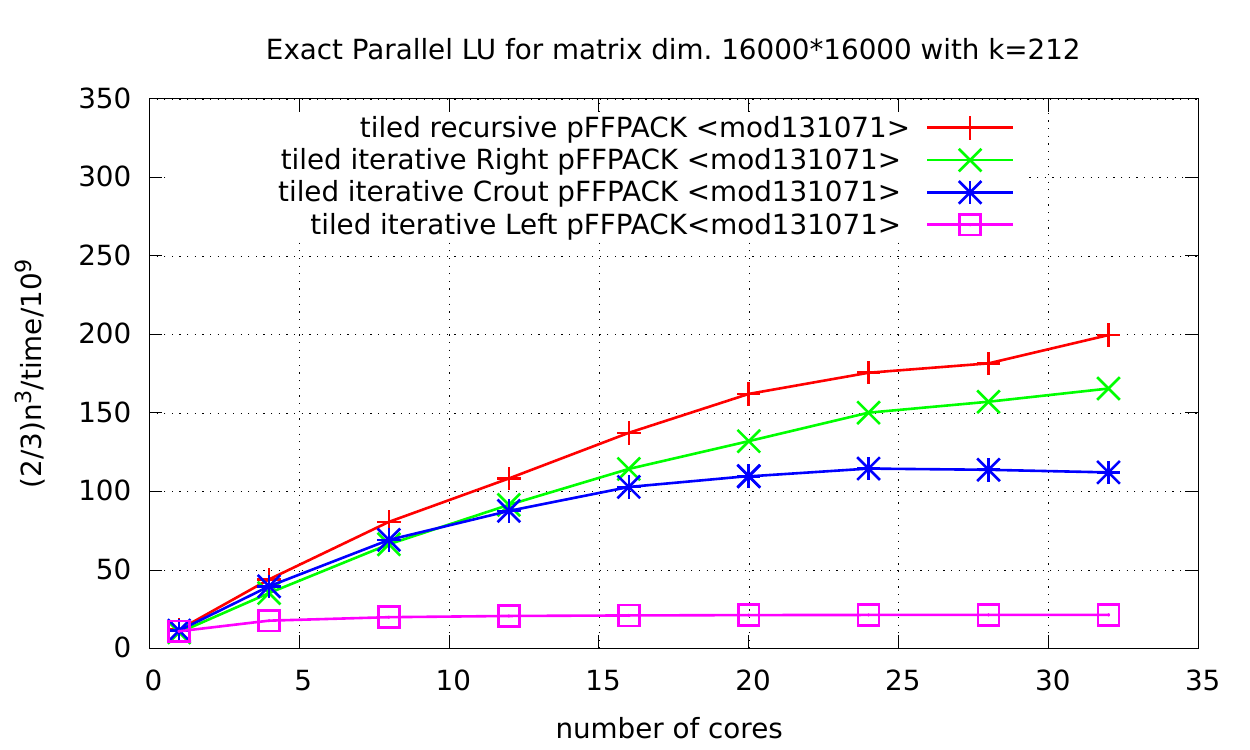}
\caption{Parallel LU factorization on full rank matrices with modular operations}
\label{fig:moduloBlockLU}
\end{figure}
In Figure~\ref{fig:moduloBlockLU} we compare the tiled iterative variants with
the tiled recursive algorithm. The latter uses as a base case an iterative Crout
algorithm too which performs fewer modular operations,

The tiled recursive algorithm performs better than all other tiled iterative
versions.
This can be explained by a finer and more adaptive  granularity and a better locality.
 
The left looking variant performs poorly for it uses an expensive sequential
\trsm task. Although Crout and right-looking variant perform about the same
number of matrix products, those of an iteration of the right-looking variant
are independent, contrarily to those of the Crout variant, which explains a
better performance despite a larger number of modular reductions.

\begin{figure}[ht!]
\centering
\includegraphics[width=.8\textwidth,angle=0.]{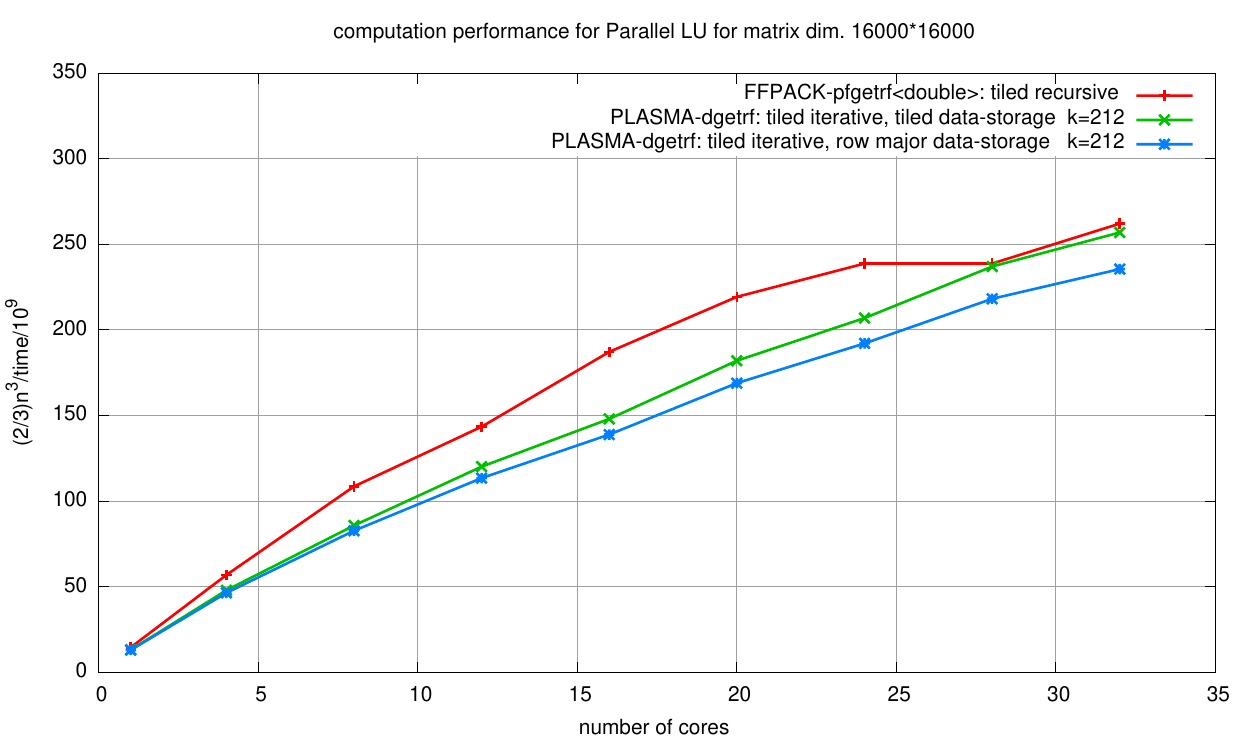}
\caption{Parallel LU factorization on full rank matrices without modular operations}
\label{fig:doubleBlockLU}
\end{figure}
Figure~\ref{fig:doubleBlockLU} shows the performance without modular
reductions, of the tiled recursive parallel implementation on full rank matrices
compared to \plasmaquark. 
The best block size for the latter library was
determined by hand for each matrix size. The two possible data-storage for
\plasmaquark are used: the collection of tiles or the row-major data-storage.
Our tiled recursive parallel PLUQ implementation without modular reductions
behaves better than the \plasmaquark \dgetrftile. This is mainly due
to the bi-dimensional cutting which allows for a faster panel elimination,  parallel \trsm
computations, more balanced \gemm computations and some use of
Strassen-Winograd's algorithm. This explains why performance join again on more
than 24 cores: the size of the sequential blocks get below the threshold where
this algorithm speeds up computations (typically 2400 on this machine).

\section{Elimination with rank deficiencies}
\label{sec:rankdef}
 
\subsection{Pivoting strategies}
We now consider the general case of matrices with arbitrary rank
profile, that can lead to rank deficiencies in the panel eliminations.

Algorithms computing the row rank profile (or equivalently the column
echelon form) used to share a common pivoting strategy:
 
to search for pivots in a row-major fashion and consider the next row only if no
non-zero pivot was found (see~\cite{JPS13} and references therein).
Such an iterative algorithm can be translated into a slab recursive algorithm
splitting the row dimension in halves (as implemented in sequential
in~\cite{DGP08}) or into a slab iterative algorithm.

More recently, we presented in~\cite{DPS13} a more flexible pivoting strategy
that results in a tile recursive algorithm, cutting both dimensions
simultaneously. As a by product, both row and column rank profiles are
also computed simultaneously.

\vspace{-1em}
\paragraph{A slab iterative algorithm.}

\begin{figure}[ht!]
\centering
\includegraphics[width=.7\textwidth,angle=0.]{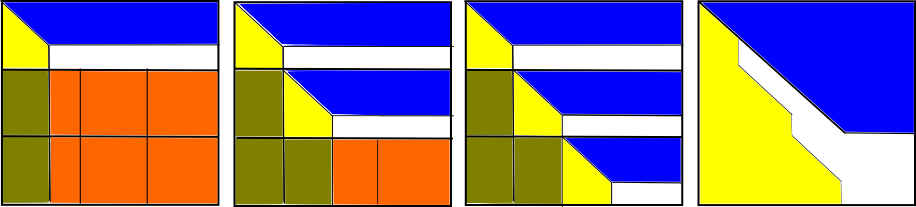}

\caption{Slab iterative factorization of a matrix with rank deficiencies}
\label{fig:TileCUP}
\end{figure}

In the slab iterative algorithm shown in Figure~\ref{fig:TileCUP}, each panel factorization has to be run by a
sequential algorithm. This sequential task is costly and therefore imposes a
choice of a fine granularity, which, as we saw, on the other hand implies more
modular reductions and a lesser speed-up of Strassen-Winograd's algorithm.

Another difficulty is the fact that the starting column position of
each panel is determined by the rank of the blocks computed so far.
It can only be determined dynamically upon the execution.
This implies in particular that no data-storage by tiles, that fit the
tiles of the algorithm is possible here. 
Moreover, the workload of each block operation may strongly vary,
depending on the rank of the corresponding slab. Such heterogeneous
tasks lead us to opt for work-stealing based runtimes instead of
static thread management.

\begin{figure}[ht!]
\centering
\includegraphics[width=0.8\textwidth,angle=0.]{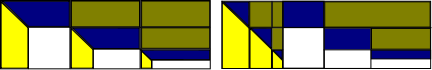}
\caption{Panel PLUQ factorization}
\label{fig:panel-pluq}
\end{figure}

\vspace{-1em}
\paragraph{Tiled iterative elimination.}
In order to speed-up the panel computation, we can split it into column
tiles. Thanks to the pivoting strategy of~\cite{DPS13}, it is still possible to
recover the rank profiles afterwards. Now with this splitting, the operations
remain more local and updates can be parallelized.
This approach shares similarities with the recursive computation of the panel
described in~\cite{DFLL11}.
Figure~\ref{fig:panel-pluq} illustrates this tile iterative factorization
obtained by the combination of a row-slab iterative algorithm, and a column-slab
iterative panel factorization.

This optimization used in the computation of the slab factorization
improved the computation speed by a factor of 2, to achieve a speed-up
of 6.5 on 32 cores with \libkomp.  

\vspace{-1em}
\paragraph{Tiled recursive elimination.}
Lastly, the tile recursive algorithm described in~\cite{DPS13} can be run in
parallel using recursive tasks and the \pfgemm, \pftrsm and \pflaswp routines.
Contrarily to most recursive factorization algorithms, the recursive splitting
is done in four quadrants.

It has the interesting feature that if the tile top-left tile is rank deficient,
then the elmination of the bottom-left and top-right tiles can be parallelized.

\begin{figure}[ht!]
\centering
\includegraphics[width=.8\textwidth,angle=0.]{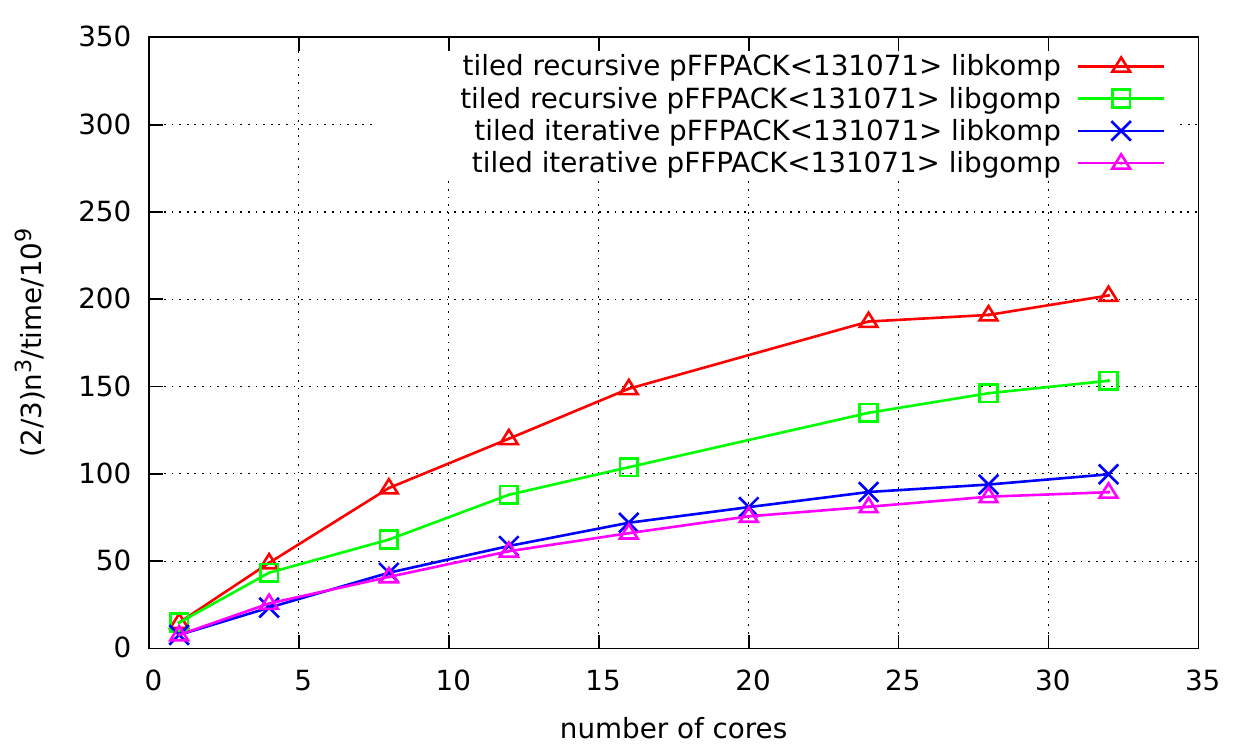}\\
\caption{Performance of tiled recursive and tiled iterative factorizations
  using  \libgomp and \libkomp. Matrix dimension $n=16000$ with rank $15500$}
\label{fig:PLUQ-Gflops}
\end{figure}

Figure~\ref{fig:PLUQ-Gflops} shows performance obtained for the tiled recursive
and the tiled iterative factorization. Both
versions are tested using \libgomp and \libkomp libraries. The input \texttt{S16K} is a
$16000\times 16000$ matrix with low rank deficiency (rank is 15500). Linearly
independant rows and columns of the generated matrix are uniformly distributed
on the dimension.

The implementation with OpenMP of the tiled recursive LU maintained high
efficiency in the case of rank deficient matrices. It attained a speed-up of
13.6 on 32 cores. Besides the fact that it benefits from Strassen-winograd
implementation, it is adapted to minimize memory accesses and optimize data
placement.

Using \libkomp instead of
\libgomp library and numactl, for round and robin interleave memory placement,
that helps reducing dependency on bus speed, we manage to obtain peak
performance for our tiled recursive LU factorization.

\section{Conclusion}
We analyzed five different algorithms for the computation of Gaussian
elimination over a finite field.
 
The granularity surely optimizes the parallelization of these
algorithms but at the cost of more modular operations. 
Algorithm optimizing modular reductions are unfortunately not the most efficient
in parallel. The best compromise is obtained with our recursive tiled algorithm
that performs best in both aspects.

\vspace{-1em}
\paragraph{Perspective}
Our future work focuses on two main issues. First, the use of specific allocators
that can be used for a better mapping of data in memory and reduce distant accesses.
Second, parallel programming frameworks for multicore processors \cite{KLDB10}
could be more effective than binding threads on each NUMA node.  
Dataflow based dependencies, like when using OpenMP 4.0 directives, 
can ensure more parallelism for recursive implementation using \libkomp\cite{BGD12} library.

\vspace{-1em}
{\scriptsize
\bibliographystyle{abbrvurl}
\bibliography{parallelPLUQ}
}

\end{document}